\documentclass[12pt]{article}
\usepackage{amssymb}
\usepackage{indentfirst}

\newcommand{\bee}{\begin{eqnarray}}
\newcommand{\ede}{\end{eqnarray}}
\begin{document}
\baselineskip=12 pt
\begin{center}
{\Large\bf Study of  the Spin-weighted Spheroidal Wave Equation in
the Case of $s=3/2$ }
\end{center}
\begin{center}
Kun Dong${}$\footnote{e-mail : woailiuyanbin1@126.com}, Guihua Tian
${}$\footnote{e-mail : hua2007@126.com}
\end{center}
\begin{center}
School of Sciences,\\
Beijing University of Posts and Telecommunications,\\
Beijing, China, 100876.\\
\end{center}
\begin{abstract}
  In this thesis we use the means of super-symmetric quantum mechanics to study of the Spin-weighted Spheroidal Wave in the case of s=3/2. We obtain some interesting results: the first-five terms of the super-potential, the general form of the super-potential. The ground eigen-function and eigenvalue of the equation are also given. According these results?we make use of the shape invariance property to compute the exited eigenvalues and eigen-functions. These results help us to understand the Spin-weighted Spheroidal Wave and show that it is integral.
\end{abstract}
PACS:11.30Pb; 04.25Nx; 04.70-s

\section{Introduction}
The spin-weighted spheroidal functions first appeared in the study
of the stable problem of Kerr black hole. Now, the spin-weighted
spheroidal functions has been widely used in many fields, such as
gravitational wave detection, quantum filed theory in curved
space-time,black hole stable problem;nuclear modeling; spheroidal
cavity problem, spheroidal electromagnetic diffraction,scattering
and similar problems in acoustic science,etc. Their  equation is
\begin{equation}
 \left[\frac{1}{\sin
\theta}\frac{d}{d\theta}\left(\sin \theta \frac{d}{d
\theta}\right)+s+ \beta^2\cos ^2 \theta -2\beta \cos \theta
-\frac{\left(m+s\cos \theta\right)^{2}}{\sin ^2
\theta}+\lambda\right]\Theta(\theta)=0,\label{original eq}
\end{equation}
The parameter $s$, the spin-weight of the perturbation fields could
be $s=0, \pm\frac12, \pm\frac32, \pm1, \pm2$, and corresponds to the
scalar, neutrino, Rarita-schwinger fields, electromagnetic or
gravitational perturbations respectively.When $\beta=0$ they reduce
to the spin-weighted spherical equations,whose solutions are the
well-known the spin-weighted spherical harmonics. Furthermore, when
$\beta=s=0$, they become to the spherical ones, whose solutions are
the famous associated Legendre's functions $P_l^m$. However when
$\beta\neq0$ and $s\neq0$, the solutions are the spin-weighted
spheroidal functions, which are treated recently by a new method in
supersymmetry quantum mechanics. Some nice results are obtained in
the case of $s=0$ and $s=\frac12$. Now we continue this study in the
case of $\frac32$. Though the method is somehow identical, the
computation involved becomes more complex than before. The course of
our calculations could provide useful information for studying them
in the case of $s=1$ and $s=2$.

\section{Calculation of super-potential and ground eigenvalue}
The boundary conditions for Eq.(\ref{original eq}) requires $\Theta$
is finite at $\theta=0,\ \pi$, this is kind of the Sturm-liuvelle
problem. As done before, Eq.(\ref{original eq}) is transformed in
the Schr\"{o}dinger form for the use of the method of supersymmetry
quantum mechanics. Changing the eigenfunction $\Theta$ by
\begin{eqnarray}
\Theta(\theta)=\frac{\Psi(\theta)}{\sqrt{sin\theta}},
\end{eqnarray}
the differential equations for the new depend functions $\Psi$
become
\begin{eqnarray}
\frac{d^2\Psi}{d\theta^2}+\left[\frac{1}{4}+s+\beta^{2}\cos ^2
\theta -2s\beta \cos\theta-\frac{(m+s\cos\theta)^2-\frac{1}{4}}{\sin
^2\theta}+E\right]\Psi=0. \label{new eq}
\end{eqnarray}
The corresponding  boundary conditions now turn out as
$\Psi|_{\theta=0}=\Psi|_{\theta=\pi}=0$. The potential in
Eq.(\ref{new eq}) is
\begin{eqnarray}
&&V(\theta,\beta,s)=-\left[\frac{1}{4}+s+\beta^{2}\cos ^2 \theta
-2s\beta \cos\theta-\frac{(m+s\cos\theta)^2-\frac{1}{4}}{\sin
^2\theta}\right]\label{potential},
\end{eqnarray}
Now we will introduce SUSYQM to our problem. According to the theory
of the $SUSYQM$, the form of ground eigenfunction  $\Psi_0$ is
completely known through the super-potential $W$ by the formula
\begin{eqnarray}
\Psi_0&=&N\exp\left[-\int Wd\theta\right]\label{psi}.
\end{eqnarray}
The problem is transformed into solving the super potential $W$. The
super potential $W$ is a most important notion in SUSYQM, and it is
determined by the $V(\theta,\beta,s)$ through the Reccita's equation
\begin{equation}
W^2-W'=V(\theta,\beta,s)-E_0.\label{potential and super relation}
\end{equation}
SO,  the key work is to solve Eq.(\ref{potential and super
relation}). As done before, by  expanding the super-potential $W$
and the ground eigenvalue $E_0$ into series form of  the parameter
$\beta$:
\begin{equation}
W=\sum_{n=0}^{\infty}\beta^nW_{n}\label{super-potential expansion}
\end{equation}
\begin{eqnarray}
E_0=\sum_{n=0}^{\infty}E_{0,n;m}\beta^n\label{energy expansion},
\end{eqnarray}
 we  solve Eq.(\ref{potential and super relation}). The results are:
\begin{eqnarray}
W'_0-W_0^2&=&E_{0,0;m}+\frac{3}{2}+\frac14-\frac{(m+\frac{3}{2}\cos\theta)^2-\frac{1}{4}}{\sin ^2\theta}\equiv f_0(\theta)\label{w0equation}\\
W'_1-2W_0W_1&=&E_{0,1;m}-3\cos\theta\equiv f_1(\theta)\\
W'_2-2W_0W_2&=&E_{0,2;m}+\cos^2\theta+W_1^2\equiv f_2(\theta)\\
W'_{n}-2W_{0}W_{n}&=&E_{0,n;m}+\sum_{k=1}^{n-1}W_{k}W_{n-k}\equiv
f_n(\theta) (n\geq3)\label{fn}
\end{eqnarray}
The solution of Eq.(\ref{w0equation}) is easy to find
\begin{eqnarray}
E_{0,0;m}&=&m^2+m-15/4\nonumber\\
W_0&=&-\frac{3/2+(m+\frac12)\cos\theta}{sin\theta}\label{W_0}
\end{eqnarray}
With $W_0$ known, it is easy to give $W_n$ on according to the
knowledge of differential equations,
\begin{eqnarray}
W_{n}(\theta)&=&e^{2\int
W_{0}d\theta}A_{n}(\theta)=\bigg[\tan\frac{\theta}2
\sin^{2m+1}\theta\bigg]^{-3} A_{n}(\theta) \label{A-n,W-n relation}
\end{eqnarray}
where,
\begin{eqnarray}
A_{n}(\theta)&=&\int f_n(\theta)e^{-2\int W_{0}d\theta} d\theta \nonumber\\
&=&\int f_n(\theta)\bigg[\tan\frac{\theta}{2}\sin^{2m+1}\theta\bigg]^{3} d\theta \nonumber\\
&=&\int f_n(\theta)(1-cos\theta)^{3}sin^{2m-2}\theta d\theta
\label{A-n eqn}.
\end{eqnarray}
Because there  appears the cubic power or inverse
 cubic powers $\tan\frac{\theta}{2}=\frac{1-\cos\theta}{\sin\theta}=\frac{\sin\theta}{1+\cos\theta}$ terms
 in Eqs.(\ref{A-n,W-n relation}),(\ref{A-n eqn}), the subsequent
 calculation becomes more complicated than before (that is,
the cases of s=0 and s=1/2). We put it in appendix A. the obtained
results are
  \begin{eqnarray}
  E_{0,1;m}&=&-\frac{9}{2m+2}\nonumber\\
E_{0,2;m}&=&-\frac{8m^{3}+96m^{2}+168m-1}{(2m+2)^{3}(2m+3)}\nonumber\\
E_{0,3;m}&=&-\frac{36(2m-1)^{2}(2m+5)^{2}}{(2m+2)^{5}(2m+3)(2m+4)}\label{E-1}\end{eqnarray}
\begin{eqnarray}
E_{0,4;m}=-\frac{4(2m-1)^{2}(2m+5)(16m^5+672m^4+3320m^3+2416m^2-6975m-7942)}{(2m+2)^{7}(2m+4)^{2}(2m+3)^{3}}\label{E-4}
\end{eqnarray}
\begin{eqnarray}
W_{1}(\theta)&=&a_{1,1}\sin\theta\\
W_{2}(\theta)&=&b_{2,1}sin\theta+a_{2,1}sin\theta\cos\theta\\
W_{3}(\theta)&=&b_{3,1}\sin\theta+b_{3,2}\sin^{3}\theta+a_{3,1}\sin\theta\cos\theta\\
W_{4}(\theta)&=&b_{4,1}\sin\theta+b_{4,1}\sin^{3}\theta+a_{4,1}\sin\theta\cos\theta+a_{4,2}\sin^{3}\theta\cos\theta
\end{eqnarray}
Where,
\begin{eqnarray}
a_{1,1}&=&-\frac{3}{2m+2}\\
b_{2,1}&=&-\frac{3(2m-1)(2m+5)}{(2m+2)^{3}(2m+3)},\\
a_{2,1}&=&\frac{(2m-1)(2m+5)}{(2m+2)^{2}(2m+3)}\\
b_{3,1}&=&\frac{108(2m-1)(2m+5)}{(2m+2)^{5}(2m+3)(2m+4)},\\
b_{3,2}&=&\frac{6(2m-1)(2m+5)}{(2m+2)^{3}(2m+3)(2m+4)}
\\
a_{3,1}&=&-\frac{36(2m-1)(2m+5)}{(2m+2)^{4}(2m+3)(2m+4)}
\\
b_{4,1}&=&\frac{36(2m-1)(16m^5+672m^4+3320m^3+2416m^2-6975m-7942)}{(2m+2)^{6}(2m+4)^{2}(2m+3)}
\\
b_{4,2}&=&-\frac{2(2m-1)(4m^3+16m^2+25m+64)}{(2m+2)^{4}(2m+4)^{2}(2m+3)}
\\
a_{4,1}&=&\frac{12(2m-1)(16m^5+672m^4+3320m^3+2416m^2-6975m-7942)}{(2m+2)^{5}(2m+4)^{2}(2m+3)}
\\
a_{4,2}&=&-\frac{4(2m-1)(4m^3+16m^2+25m+64)}{(2m+2)^{4}(2m+4)^{2}(2m+3)^{2}}
\end{eqnarray}

\section{Summarize and prove the general formula of super-potential }
 From the four terms of $W_1-W_4$, we hypothetically summarize a
general formula for $W_n$ as
\begin{equation}
W_{n}(\theta)=\sum_{k=1}^{[\frac{n}{2}]}a_{n,k}\sin^{2k-1}\theta\cos\theta+\sum_{k=1}^{[\frac{n+1}{2}]}b_{n,k}\sin^{2k-1}\theta\label{w_n}
\end{equation}
Here we use mathematical induction to prove that the guess is true.

First it is easy to see the assumption (\ref{w_n}) is the same as
that of $W_1$ when $N=1$. Under the condition that all $W_N$ meet
the requirement of (\ref{w_n}) whenever $N\le n-1$, we will try to
solve the differential equation for $W_n$ to verify that it also can
be written as that of (\ref{w_n}) and be determined by the terms
$W_k, k<n$. The results are
\begin{eqnarray}
&&b_{n,l}\nonumber\\
&=&\sum_{p=l+1}^{\left[\frac{n}{2}\right]+1}\frac{3(2m+2l-2)j_{n,p}\bar{I}(2m+2p-4,p-l-1)}{(2m+2l+1)(2m+2l+3)(2m+2p-3)}+\frac{g_{n,l}}{2m+2l}
\delta_{l,1} ,l\ge1 \label{bl}
\end{eqnarray}
\begin{eqnarray}
&&a_{n,l}\nonumber\\
&=&-\sum_{p=l+1}^{\left[\frac{n}{2}\right]+1}\frac{(2m+2l-2)(2m+2l)j_{n,p}\bar{I}(2m+2p-4,p-l-1)}{(2m+2l+1)(2m+2l+3)(2m+2p-3)},l\ge1
\label{al}
\end{eqnarray}
where\begin{eqnarray} j_{n,p}&=&
\frac{[h_{n,p}(2m+2p)-3g_{n,p}](2m+2p+1)}{(2m+2p-2)(2m+2p)}.
\label{j-np}
\end{eqnarray}
The terms $h_{n,p},\ g_{n,p}$ are determined by the coefficients
$a_{k,j},\ b_{k,j}, \ k<n$ of $W_k,\ k<n$:
\begin{eqnarray}
h_{n,p}&=&\sum_{k=1}^{n-1}\sum_{j=1}^{p-1}\bigg[b_{k,p-j}b_{n-k,j}+a_{k,p-j}a_{n-k,j}
-a_{k,p-1-j}a_{n-k,j}\bigg]\\
 g_{n,p}&=&\sum_{k=1}^{n-1}\sum_{j=1}^{p-1}\bigg[b_{k,p-j}a_{n-k,j}+a_{k,p-j}b_{n-k,j}\bigg]\end{eqnarray}
\section{The ground eigenfunction }
Finally, according to the shape invariant potential,we obtain the
wave functions of the exited state $\Theta_0$.
\begin{eqnarray}
W&=&W_0+\sum_{n=1}^{\infty}\left[\cos\theta\sum_{k=1}^{[\frac{n}{2}]}a_{n,k}\sin^{2k-1}\theta+\sum_{k=1}^{[\frac{n+1}{2}]}b_{n,k}\sin^{2k-1}\theta\right]\beta^n
\end{eqnarray}
The ground eigenfunction becomes
\begin{eqnarray}
\Theta_0&=&N\exp\left[-\int Wd\theta\right]\nonumber\\
        &=&N\exp\left[-\int W_0d\theta-\sum_{n=1}^{\infty}\int W_nd\theta\right]\nonumber\\
        &=&N\exp\left[\int \frac{3/2+(m+\frac12)\cos\theta}{sin\theta} d\theta-\beta^n\sum_{n=1}^{\infty}
        \int (\cos\theta\sum_{k=1}^{[\frac{n}{2}]}a_{n,k}\sin^{2k-1}\theta +\sum_{k=1}^{[\frac{n+1}{2}]}b_{n,k}\sin^{2k-1}\theta )d\theta\right] \nonumber\\
        &=&N(1-\cos\theta)^{\frac{3}{2}}\sin^{m-1}\theta\exp\left[-\beta^n\sum_{n=1}^{\infty}
        \left(\sum_{k=1}^{[\frac{n}{2}]}a_{n,k}\frac{\sin^{2k}\theta}{2k}+\sum_{k=1}^{[\frac{n+1}{2}]}b_{n,k}P(2k-1,\theta)\right)\right]
\end{eqnarray}
The ground eigenvalue is
\begin{eqnarray}
E_{0,n;m}&=&E_{0,0;m}+\sum_{n=1}^{\infty}\beta^nE_{0,n;m}
=m^2+m-15/2+\sum_{n=1}^{\infty}\beta^nE_{0,n;m}
\end{eqnarray}

\section{The excited eigenfunctions  }
In the following, we will compute the excited eigenfunctions. As
done in Ref.\cite{Tian12},  we hope to extend the study of the
recurrence relations by the means of super-symmetric quantum
mechanics to Eq. (\ref{new eq}).

The super-potential $W$  connects the two partner potential
$V_{\mp}$ by
\begin{equation} \label{vpm} V^{\mp}(\theta)=W^2(\theta) \mp W'(\theta).
\end{equation}
The shape-invariance properties mean that the pair of partner
potentials $V^{\pm}(x)$  are similar in shape and differ only in the
parameters, that is
\begin{equation} \label{shape invariant} V^+(\theta;a_1) = V^-(\theta;a_2) +
R(a_1),
\end{equation}
where $a_1$ is a set of parameters, $a_2$ is a function of $a_1$
(say $a_2=f(a_1)$) and the remainder $R(a_1)$ is independent of
$\theta$.

We must introduce the parameters $A_{i,j},\ B_{i,j}$ into the
super-potential $W$ in order to study the shape-invariance
properties of the spin-weighted spheroidal equations as:
\begin{eqnarray}
W(A_{n,j},B_{n,j},\theta)&=&-A_{0,0}(m+\frac12)\cot
\theta-\frac{3}{2}B_{0,0}\csc\theta \nonumber\\ &&
+\sum_{n=1}^{\infty}\beta^nW_n(A_{n,j},B_{n,j},\theta),
\end{eqnarray}
where
\begin{eqnarray}
&&W_n(A_{n,j},B_{n,j},\theta)\nonumber\\
&& \ \ \ \
=\sum_{j=1}^{[\frac{n+1}{2}]}\bar{b}_{n,j}\sin^{2j-1}\theta +\cos
\theta\sum_{j=1}^{[\frac{n}{2}]}\bar{a}_{n,j}\sin^{2j-1}\theta
\end{eqnarray}
with
\begin{eqnarray}
\bar{a}_{n,j} =A_{n,j}a_{n,j},\  \bar{b}_{n,j} =B_{n,j}b_{n,j}
\end{eqnarray}
Then, $V^{\pm}(A_{n,j},B_{n,j},\theta)$ are
$V^{\pm}(A_{n,j},B_{n,j},\theta)$ are defined as
\begin{eqnarray}
V^{\pm}(A_{n,j},B_{n,j},\theta)=W^2(A_{n,j},B_{n,j},\theta)\pm W'\nonumber\\
=\sum_{n=0}^{\infty}\beta^nV^{\pm}_n(A_{i,j},B_{n,j},\theta).
\end{eqnarray}
The key point is to try to find some quantities $C_{i,j},D_{i,j}$ to
make the relations
\begin{eqnarray}
V^{+}_n(A_{i,j},B_{n,j},\theta)=V^{-}_n(C_{i,j},D_{n,j},\theta)+R_{n;m}(A_{i,j},B_{n,j})\label{shape-invariance
v-pm }
\end{eqnarray}
retain with  $R_{n;m}(A_{i,j},B_{n,j})=R_{n;m}$ pure quantities. In
the previous , we know the general formula with $W_n$ in the case of
$s=3/2$ is same as $s=1/2$. So that, we can  refer to the results of
previous article:
\begin{eqnarray}
D_{n,p}&=&\frac{D_{0,0}a_{n,p}}{\alpha_{p}b_{n,p}}C_{n,p}-\frac{U_{n,p}}{\alpha_{p}b_{n,p}}\label{d-np}\\
C_{n,p-1}&=&
\frac{\bigg(\alpha_{p}+\frac{D^2_{0,0}}{\alpha_{p}}\bigg)a_{n,p}}{(\alpha_{p}-1)a_{n,p-1}}C_{n,p}+
\frac{\check{U}_{n,p}-\frac{D_{0,0}}{\alpha_{p}}U_{n,p}}
{(\alpha_{p}-1)a_{n,p-1}},\label{c-np}\\
&& \ \ \ \ \ \ \ \ \ p=2,3,\ldots, [\frac{n+2}2]
\end{eqnarray}
where,
\begin{eqnarray}
U_{n,p}&=&-\alpha_{p}D_{n,p}b_{n,p}-D_{0,0}C_{n,p}a_{n,p} \\
\check{U}_{n,p}&=&-\alpha_{p}C_{n,p}a_{n,p}-D_{0,0}D_{n,p}b_{n,p} +
(\alpha_{p}-1)C_{n,p-1}a_{n,p-1}\\
\alpha_{p}&=&\bigg[(2m+1)C_{0,0}+(2p-1)\bigg]
\end{eqnarray}
The only difference is the initial value. Now, we will give the
initial value for the cases $n=0,\ 1,2$:
\begin{eqnarray}
C_{0,0}&=&A_{0,0}+\frac2{2m+1},\\ D_{0,0}&=&B_{0,0}\\
D_{1,1}&=&\frac{(2m+1)A_{0,0}-1}{(2m+1)A_{0,0}+3}B_{1,1}
\end{eqnarray} and
\begin{eqnarray}
D_{2,1}&=&\frac{(2m+1)A_{0,0}-1}{(2m+1)A_{0,0}+3}B_{2,1}\nonumber\\ &&+\frac{18B_{0,0}B_{2,1}}{[(2m+1)A_{0,0}+3][(2m+1)A_{0,0}+4]}\nonumber\\
&&-\frac{24[(2m+1)A_{0,0}+1]B_{0,0}B_{1,1}^{2}}{[(2m+1)A_{0,0}+3]^{3}[(2m+1)A_{0,0}+4]}
\\
C_{2,1}&=&\frac{8[(2m+1)A_{0,0}+1]B_{1,1}^{2}}{[(2m+1)A_{0,0}+3]^{3}[(2m+1)A_{0,0}+4]}\nonumber\\
       &&+\frac{(2m+1)A_{0,0}-2}{(2m+1)A_{0,0}+4}A_{2,1}
\end{eqnarray}
with
\begin{eqnarray}
&&R_{0;m}(A_{0,0})=(2m+1)A_{0,0}+1,\\
&&R_{1;m}(A_{0,0},B_{0,0},B_{1,1})=-\frac{12B_{0,0}B_{1,1}}{(2m+1)A_{0,0}+3}
\\
&&R_{2;m}(A_{0,0},B_{0,0},B_{1,1},B_{2,1},A_{2,1})\nonumber\\
&=&[-\frac{4B_{0,0}B_{2,1}}{(2m+1)A_{0,0}+3}+AB_{1,1}^{2}+BA_{2,1}]
\end{eqnarray}
where
\begin{eqnarray}
A=\frac{72B_{0,0}^{2}-8[(2m+1)A_{0,0}-1][(2m+1)A_{0,0}+3]}{[(2m+1)A_{0,0}+3]^{3}[(2m+1)A_{0,0}+4]}\nonumber\\
B=\frac{54B_{0,0}^{2}-2[(2m+1)A_{0,0}-1][(2m+1)A_{0,0}+3]}{[(2m+1)A_{0,0}+3][(2m+1)A_{0,0}+4]}
\end{eqnarray}

 Then, the  excited eigen-values $E_{l;m}$ and
eigenfunctions $\Psi_l$ is obtained by the recurrence relation :
\begin{eqnarray}
&&  E_{l;m}^-=E_{0;m}+\sum_{k=1}^{l}R(a_{k},b_{k}),\\&&
E_{0;m}=m(m+1)-\frac{15}{2}+\sum_{n=1}^{\infty}E_{0,n;m}\beta^n
\\
&&
R(a_{k},b_{k})=R_{0;m}+\sum_{n=1}^{\infty}\beta^nR_{n;m}(a_{k},b_{k}),\\
&&a_1=(A_{i,j},B_{i,j}),\ a_2=(C_{i,j},D_{i,j}),\ldots, \\ &&
\Psi_0\propto
\exp\bigg[-\int_{\theta_0}^{\theta}W(A_{n,j},B_{n,j},\theta)d\theta\bigg],
\\ && {\cal A}^{\dagger}=-\frac d{d\theta}+W(A_{n,j},B_{n,j},\theta)\\
&& \Psi_n^-={\cal A}^{\dagger}
\,(A_{n,j},B_{n,j},\theta)\Psi_{n-1}^-(C_{n,j},D_{n,j},\theta),\\ &&
\  \ \ \ \ \ \ \ \ \ \  \ n=1,2,3,\dots
\end{eqnarray}
In conclusion, we have proved the shape-invariance properties for
the spin-weighted equations in the case of $s=\frac32$ and obtain
the recurrence relations for them. By these results we can get the
exited eigenvalue and eigenfunction.

\section*{Acknowledgements}
 This work is supported by the National Natural Science Foundation of China (Grant Nos.10875018,10773002)
\section{The details of the calculation of the first several terms of the superpotential}

In order to calculate Eqs.(\ref{A-n,W-n relation})-(\ref{A-n eqn}),
we needs the following integral formulae \cite{grad}:
\begin{eqnarray}
P(2m,\theta)=\int{sin^{2m}\theta}d\theta=-\frac{\cot\theta}{2m+1}\left[\sum_{k=0}^{m-1}\bar{I}(2m,k)sin^{2m-2k}\theta\right]+\frac{(2m-1)!!}{(2m)!!}\label{defination
P}\label{p2m}
\end{eqnarray}
where
\begin{eqnarray}
\bar{I}(2m,k)=\frac{(2m+1)(2m-1)\cdot\cdot\cdot(2m-2k+1)}{2m(2m-2)\cdot\cdot\cdot(2m-2k)},(k\geq0),
\end{eqnarray}
the repeating results of the above equation
\begin{eqnarray}
P(2m,\theta)&=&\frac{2m-1}{2m}P(2m-2,\theta)-\frac{\cos\theta\sin^{2m-1}\theta}{2m},
\end{eqnarray}
\begin{eqnarray}
P(2m+2,\theta)&=&\frac{(2m-1)(2m+1)}{2m(2m+2)}P(2m-2,\theta)\nonumber\\
&&-\frac{\cos\theta\sin^{2m+1}}{2m+2}-\frac{(2m+1)\cos\theta\sin^{2m-1}\theta}{2m(2m+2)}\nonumber\\
P(2m+4,\theta)&=&\frac{(2m-1)(2m+1)(2m+3)}{2m(2m+2)(2m-4)}P(2m-2,\theta)-\frac{\cos\theta\sin^{2m+3}}{2m+4}\nonumber\\
&&-\frac{(2m+3)\cos\theta\sin^{2m+1}\theta}{(2m+2)(2m+4)}-\frac{(2m+1)(2m+3)\cos\theta\sin^{2m-1}\theta}{2m(2m+2)(2m-4)}
\end{eqnarray}
By the above equation,we can summarize the general formula
\begin{eqnarray}
P(2m+2n-2)&=&-\cos\theta\sum_{l=1}^{n}\frac{\bar{I}(2m+2n-2,n-l)}{2m+2n-1}\sin^{2m+2l-3}\theta\nonumber\\
&&
+\frac{2m-1}{2m+2n-1}\bar{I}(2m+2n-2,n-1)P(2m-2,\theta)\label{Integral
formula}
\end{eqnarray}
This formula can be proved by mathematical induction. By the help of
Eqs.(\ref{W_0}), (\ref{Integral formula}), $A_{1}(\theta)$ is now
simplified as
\begin{eqnarray}
&&A_{1}(\theta)\nonumber\\
&=&\int (E_{0,1;m}-3cos\theta)(1-cos\theta)^{3}sin^{2m-2}\theta d\theta\nonumber\\
     &=&(4E_{0,1;m}+12)P(2m-2,\theta)-(3E_{0,1;m}+15)P(2m,\theta)+3P(2m+2,\theta)\nonumber\\
     &&-\frac{4E_{0,1;m}+12)sin^{2m-1}\theta}{2m-1}-\frac{(9+E_{0,1;m})sin^{2m+1}\theta}{2m+1}\nonumber\\
     &=&[4E_{0,1;m}+12-(3E_{0,1;m}+15)\frac{2m-1}{2m}+\frac{3(2m-1)(2m+1)}{2m(2m+2)}]P(2m-2,\theta)\nonumber\\
     &&-\frac{4(E_{0,1;m}+12)sin^{2m-1}\theta}{2m-1}-\frac{(9+E_{0,1;m})sin^{2m+1}\theta}{2m+1}\nonumber\\
     &&-\frac{3\cos\theta\sin^{2m+1}}{2m+2}+\frac{[(2m+2)(3E_{0,1;m}+15)-(2m+1)]\cos\theta\sin^{2m-1}\theta}{2m(2m+2)}\label{A-1 eqn}
\end{eqnarray}
Please note that the coefficient of $P(2m-2,\theta)$,According to
Eq.(\ref{psi}) and Eq.(\ref{p2m}), we can see that $\Psi(\theta)$ is
$\infty$ at the boundaries $\theta=0,\ \pi$. This result does not
meet the boundary conditions that $\Psi(\theta)$ should finite at
$\theta=0,\ \pi$. So that the coefficient of the term
$P(2m-2,\theta)$ should be zero.
\begin{equation}
E_{0,1;m}=-\frac{9}{2m+2}\label{E-1}
\end{equation}
and it further provides the concise form for $A_1$
\begin{eqnarray}
A_{1}(\theta)&=&-\frac{12}{2m+2}\sin^{2m-1}\theta(\cos\theta-1)\nonumber\\
             &&+\frac{3}{2m+2}\sin^{2m+1}\theta(1-\cos\theta)+\frac{3}{2m+2}\sin^{2m+1}
\end{eqnarray}
With the help of Eq.(\ref{A-n,W-n relation}),  it is easy to obtain
the first order $W_1(\theta)$
\begin{eqnarray}
W_{1}(\theta)&=&A_{1}(\theta)(1+cos\theta)^3(sin\theta)^{-2m-4}=-\frac{3}{2m+2}\sin\theta
\end{eqnarray}
Now, we can see $W_{1}(\theta)$ is convergence at the boundaries
$\theta=0,\ \pi$. So that, the result of $E_{0,1;m}$ is appropriate.
By the same steps with more complex calculation than $W_1(\theta)$ ,
we can also get $E_{0,2;m}$ , $W_{2}$ ¡¢$E_{0,3;m}$ , $W_{3}$ and
$E_{0,4;m}$ , $W_{4}$ .
\section{the calculation for the general form of $W_n$}
Back to Eqs.(\ref{fn}),(\ref{A-n,W-n relation}),(\ref{A-n eqn}), one
needs to simplify the term $\sum_{k=1}^{n-1}W_{k}W_{n-k}$ in order
to calculate $W_n$. Whenever $1\le k \le n-1$, one has $1\le n-k\le
n-1$ and $W_k(\theta)$, $W_{n-k}(\theta)$ could be written in the
form of (\ref{w_n}). That is,
\begin{eqnarray}
W_{k}(\theta)&=&\sum_{i=1}^{[\frac{k}{2}]}a_{k,i}\sin^{2i-1}\theta\cos\theta+\sum_{i=1}^{[\frac{k+1}{2}]}b_{k,i}\sin^{2i-1}\theta\nonumber\\
W_{n-k}(\theta)&=&\sum_{j=1}^{[\frac{n-k}{2}]}a_{n-k,j}\sin^{2j-1}\theta\cos\theta+\sum_{j=1}^{[\frac{n-k+1}{2}]}b_{n-k,j}\sin^{2j-1}\theta
\end{eqnarray}
For the sake of later use, the facts are true
\begin{eqnarray}
a_{i,j}=0, j<1 \ or\  j>[\frac{i}2];\ \ b_{i,j}=0, \ j<1  \ or\
j>[\frac{i+1}2]\label{a-[i,j]=0 condition}
\end{eqnarray}
for $i<n$.  Thus
\begin{eqnarray}
&&\sum_{k=1}^{n-1}W_{k}W_{n-k}\nonumber\\
&=&\sum_{k=1}^{n-1}\sum_{i=1}^{[\frac{k+1}{2}]}\sum_{j=1}^{[\frac{n-k+1}{2}]}b_{k,i}b_{n-k,j}\sin^{2i+2j-2}\theta
+\sum_{k=1}^{n-1}\sum_{i=1}^{[\frac{k}{2}]}\sum_{j=1}^{[\frac{n-k}{2}]}a_{k,i}b_{n-k,j}\sin^{2i+2j-2}\theta\cos^{2}\theta\nonumber\\
&&+\cos\theta\sum_{k=1}^{n-1}\bigg[\sum_{i=1}^{[\frac{k}{2}]}\sum_{j=1}^{[\frac{n-k+1}{2}]}a_{k,i}b_{n-k,j}\sin^{2i+2j-2}\theta
+\sum_{i=1}^{[\frac{k+1}{2}]}\sum_{j=1}^{[\frac{n-k}{2}]}b_{k,i}a_{n-k,j}\sin^{2i+2j-2}\theta\bigg]\\
&=&\sum_{p=2}^{\bar{c}_1}\sum_{k=1}^{n-1}\sum_{j=1}^{p-1}b_{k,p-j}b_{n-k,j}sin^{2p-2}\theta
+\sum_{p=2}^{\bar{c}_2}\sum_{k=1}^{n-1}\sum_{j=1}^{p-1}a_{k,p-j}a_{n-k,j}sin^{2p-2}\theta\cos^{2}\theta\nonumber\\
&&+\cos\theta\bigg[\sum_{p=2}^{\bar{c}_3}\sum_{k=1}^{n-1}\sum_{j=1}^{p-1}b_{k,p-j}a_{n-k,j}sin^{2p+1}\theta
+\sum_{p=2}^{\bar{c}_4}\sum_{k=1}^{n-1}\sum_{j=1}^{p-1}a_{k,p-j}b_{n-k,j}sin^{2p+1}\theta\bigg]\nonumber\\
&=&\sum_{p=2}^{\left[\frac{n}{2}\right]+1}\bigg[h_{n,p}+g_{n,p}\cos\theta\bigg]sin^{2p-2}\theta
\end{eqnarray}
where  $g_{n,p}$ and $h_{n,p}$ are constant coefficients:
\begin{eqnarray}
h_{n,p}&=&\sum_{k=1}^{n-1}\sum_{j=1}^{p-1}\bigg[b_{k,p-j}b_{n-k,j}+a_{k,p-j}a_{n-k,j}
-a_{k,p-1-j}a_{n-k,j}\bigg]\\
 g_{n,p}&=&\sum_{k=1}^{n-1}\sum_{j=1}^{p-1}\bigg[b_{k,p-j}a_{n-k,j}+a_{k,p-j}b_{n-k,j}\bigg]\\
 \bar{c}_1&=& \left[\frac{k+1}{2}\right]+\left[\frac{n-k+1}{2}\right],\ \bar{c}_2=\left[\frac{k}{2}\right]+\left[\frac{n-k}{2}\right]\\
 \bar{c}_3&=& \left[\frac{k+1}{2}\right]+\left[\frac{n-k}{2}\right],\ \bar{c}_4=\left[\frac{k}{2}\right]+\left[\frac{n-k+1}{2}\right]
\end{eqnarray}
It is easy to see \begin{eqnarray}
 \bar{c}_1&=&\frac n2+1,\  \ \bar{c}_2=\bar{c}_3=\bar{c}_4=\frac n2, \ when\  n\  is\  even \\
 \ \bar{c}_2&=&\frac{n-1}2,\ \ \bar{c}_1= \bar{c}_3=
 \bar{c}_4=\frac{n+1}2,\ \ when\  n\  is\ odd
\end{eqnarray}
Hence, one has
\begin{eqnarray}
 g_{n,p}&=&0,\  \ p<1\ or p> \frac n2;\ \  h_{n,p}=0,\  \ p<1\ or\  p> \frac n2+1, \ when\  n\  is\  even \nonumber\\
 g_{n,p}&=&h_{n,p}=0,\  \ p<1\ or\ p> \frac{n+1}2,\ \ when\  n\  is\
 odd \label{g-np,h-np=0 condition}.
\end{eqnarray}
We have  used  the fact (\ref{a-[i,j]=0 condition}) and  substituted
the quantities $\bar{c}_1,\ \bar{c}_2,\ \bar{c}_3,\ \bar{c}_4$ by
the maximum $\left[\frac{n}{2}\right]+1$ of them in last line in the
above equation. Taking
$f_n(z)=E_{0,n;m}+\sum_{k=1}^{n-1}W_{k}W_{n-k}$ into Eqs.(\ref{A-n
eqn}) we can have
\begin{eqnarray}
&&A_{n}(\theta)\nonumber\\
&=&\int[E_{0,n;m}+\sum_{p=2}^{\left[\frac{n}{2}\right]+1}\bigg[h_{n,p}+g_{n,p}\cos\theta\bigg]sin^{2p-2}\theta]
(1-cos\theta)^{3}sin^{2m-2}\theta d\theta\nonumber\\
&=&\int4E_{0,n;m}\sin^{2m-2}\theta
d\theta-\int3E_{0,n;m}\sin^{2m}\theta
d\theta\nonumber\\
&&-\frac{4E_{0,n;m}}{2m-1}\sin^{2m-1}\theta+\frac{E_{0,n;m}}{2m+1}\sin^{2m+1}\theta\nonumber\\
&&+\sum_{p=2}^{\left[\frac{n}{2}\right]+1}\frac{4g_{n,p}-4h_{n,p}}{2m+2p-3}sin^{2m+2p-3}\theta+\sum_{p=2}^{\left[\frac{n}{2}\right]+1}\frac{h_{n,p}-3g_{n,p}}{2m+2p-1}sin^{2m+2p-1}\theta\nonumber\\
&&+\sum_{p=2}^{\left[\frac{n}{2}\right]+1}\int
(4h_{n,p}-4g_{n,p})sin^{2m+2p-4}\theta
d\theta\nonumber\\
&&+\sum_{p=2}^{\left[\frac{n}{2}\right]+1}\int
(5g_{n,p}-3h_{n,p})sin^{2m+2p-2}\theta
d\theta-\sum_{p=2}^{\left[\frac{n}{2}\right]+1}\int
g_{n,p}sin^{2m+2p}\theta d\theta\nonumber\\
&=&4E_{0,n;m}P(2m-2,\theta)-3E_{0,n;m}P(2m,\theta)-\frac{4E_{0,n;m}}{2m-1}\sin^{2m-1}\theta+\frac{E_{0,n;m}}{2m+1}\sin^{2m+1}\theta\nonumber\\
&&+\sum_{p=2}^{\left[\frac{n}{2}\right]+1}\frac{4g_{n,p}-4h_{n,p}}{2m+2p-3}sin^{2m+2p-3}\theta+\sum_{p=2}^{\left[\frac{n}{2}\right]+1}\frac{h_{n,p}-3g_{n,p}}{2m+2p-1}sin^{2m+2p-1}\theta\nonumber\\
&&+\sum_{p=2}^{\left[\frac{n}{2}\right]+1}
(4h_{n,p}-4g_{n,p})P(2m+2p-4,\theta) \nonumber\\
&&+\sum_{p=2}^{\left[\frac{n}{2}\right]+1}
(5g_{n,p}-3h_{n,p})P(2m+2p-2,\theta)-\sum_{p=2}^{\left[\frac{n}{2}\right]+1}
g_{n,p}P(2m+2p,\theta)\nonumber\\
&=&4E_{0,n;m}P(2m-2,\theta)-3E_{0,n;m}P(2m,\theta)-\frac{4E_{0,n;m}}{2m-1}\sin^{2m-1}\theta+\frac{E_{0,n;m}}{2m+1}\sin^{2m+1}\theta\nonumber\\
&&+\sum_{p=2}^{\left[\frac{n}{2}\right]+1}\frac{4g_{n,p}-4h_{n,p}}{2m+2p-3}sin^{2m+2p-3}\theta+\sum_{p=2}^{\left[\frac{n}{2}\right]+1}\frac{h_{n,p}-3g_{n,p}}{2m+2p-1}sin^{2m+2p-1}\theta\nonumber\\
&&-\sum_{p=2}^{\left[\frac{n}{2}\right]+1}\left[\frac{5g_{n,p}-3h_{n,p}}{2m+2p-2}+\frac{(2m+2p-1)g_{n,p}}{(2m+2p-2)(2m+2p)}\right]sin^{2m+2p-3}\theta\cos\theta\nonumber\\
&&+\sum_{p=2}^{\left[\frac{n}{2}\right]+1}\frac{g_{n,p}}{2m+2p}sin^{2m+2p-1}\theta\cos\theta\nonumber\\
&&+\sum_{p=2}^{\left[\frac{n}{2}\right]+1}j_{n,p}P(2m+2p-4,\theta)
\nonumber\\
&=&4E_{0,n;m}P(2m-2,\theta)-3E_{0,n;m}P(2m,\theta)-\frac{4E_{0,n;m}}{2m-1}\sin^{2m-1}\theta+\frac{E_{0,n;m}}{2m+1}\sin^{2m+1}\theta\nonumber\\
&&+\sum_{p=2}^{\left[\frac{n}{2}\right]+1}\frac{4g_{n,p}-4h_{n,p}}{2m+2p-3}sin^{2m+2p-3}\theta+\sum_{p=2}^{\left[\frac{n}{2}\right]+1}\frac{h_{n,p}-3g_{n,p}}{2m+2p-1}sin^{2m+2p-1}\theta\nonumber\\
&&-\sum_{p=2}^{\left[\frac{n}{2}\right]+1}i_{n,p}sin^{2m+2p-3}\theta\cos\theta+\sum_{p=2}^{\left[\frac{n}{2}\right]+1}\frac{g_{n,p}}{2m+2p}sin^{2m+2p-1}\theta\cos\theta\nonumber\\
&&+\sum_{p=2}^{\left[\frac{n}{2}\right]+1}j_{n,p}P(2m+2p-4,\theta)
\end{eqnarray}
Where $i_{n,p}$ and $j_{n,p}$ are constant coefficients:
\begin{eqnarray}
i_{n,p}=\left[\frac{5g_{n,p}-3h_{n,p}}{2m+2p-2}+\frac{(2m+2p-1)g_{n,p}}{(2m+2p-2)(2m+2p)}\right]
\end{eqnarray}
\begin{eqnarray}
j_{n,p}&=&4(h_{n,p}-g_{n,p})+\frac{2m+2p-3}{2m+2p-2}[5g_{n,p}-3h_{n,p}]\nonumber\\
&&-\frac{(2m+2p-3)(2m+2p-1)}{(2m+2p)(2m+2p-2)}g_{n,p}\nonumber\\
&=& 4(h_{n,p}-g_{n,p})+(2m+2p-3)i_{n,p}\label{j-np
original form}\\
&=& \frac{[h_{n,p}(2m+2p)-3g_{n,p}](2m+2p+1)}{(2m+2p-2)(2m+2p)}
\label{j-np}
\end{eqnarray}
According to Eq.(\ref{Integral formula}), one has
\begin{eqnarray}
P(2m+2p-4)&=&-\cos\theta\sum_{l=1}^{p-1}\frac{\bar{I}(2m+2p-4,p-l-1)}{2m+2p-3}\sin^{2m+2l-3}\theta\nonumber\\
&&+\frac{2m-1}{2m+2p-3}\bar{I}(2m+2p-4,p-2)P(2m-2,\theta)\label{Integral
formula n=p-1}
\end{eqnarray}
Hence,
\begin{eqnarray}
&&A_{n}(\theta)\nonumber\\
&=&4E_{0,n;m}P(2m-2,\theta)+3E_{0,n;m}[\frac{cos\theta sin^{2m-1}\theta}{2m}-\frac{2m-1}{2m}P(2m-2,\theta)]\nonumber\\
&&-\frac{4E_{0,n;m}}{2m-1}\sin^{2m-1}\theta+\frac{E_{0,n;m}}{2m+1}\sin^{2m+1}\theta\nonumber\\
&&+\sum_{p=2}^{\left[\frac{n}{2}\right]+1}\frac{4g_{n,p}-4h_{n,p}}{2m+2p-3}sin^{2m+2p-3}\theta+\sum_{p=2}^{\left[\frac{n}{2}\right]+1}\frac{h_{n,p}-3g_{n,p}}{2m+2p-1}sin^{2m+2p-1}\theta\nonumber\\
&&-\sum_{p=2}^{\left[\frac{n}{2}\right]+1}i_{n,p}sin^{2m+2p-3}\theta\cos\theta+\sum_{p=2}^{\left[\frac{n}{2}\right]+1}\frac{g_{n,p}}{2m+2p}sin^{2m+2p-1}\theta\cos\theta\nonumber\\
&&+\sum_{p=2}^{\left[\frac{n}{2}\right]+1}j_{n,p}[-\cos\theta\sum_{l=1}^{p-1}\frac{\bar{I}(2m+2p-4,p-l-1)}{2m+2p-3}\sin^{2m+2l-3}\theta\nonumber\\
&&+\frac{2m-1}{2m+2p-3}\bar{I}(2m+2p-4,p-2)P(2m-2,\theta)]
\end{eqnarray}
For the coefficient of divergent term must be zero, that is
\begin{eqnarray}
b_1=\frac{(2m+3)}{2m}E_{0,n;m}+\sum_{p=2}^{\left[\frac{n}{2}\right]+1}j_{n,p}\frac{2m-1}{2m+2p-3}\bar{I}(2m+2p-4,p-2)=0\label{b-1}\end{eqnarray}

then
\begin{eqnarray}
&&E_{0,n;m}\nonumber\\
&=&-\frac{2m}{2m+3}\sum_{p=2}^{\left[\frac{n}{2}\right]+1}j_{n,p}\frac{2m-1}{2m+2p-3}\bar{I}(2m+2p-4,p-2)\nonumber\\
&=&-\sum_{p=2}^{\left[\frac{n}{2}\right]+1}\frac{2m(2m-1)[h_{n,p}(2m+2p)-3g_{n,p}](2m+2p+1)}{(2m+3)(2m+2p-3)(2m+2p-2)(2m+2p)}\bar{I}(2m+2p-4,p-2) \nonumber\\
\label{E_0,n;m}
\end{eqnarray}
Taking Eq.(\ref{E_0,n;m}) into Eq.(\ref{A-n-1}),we can get
\begin{eqnarray}
&&A_{n}(\theta)\nonumber\\
&=&-\frac{4E_{0,n;m}}{2m-1}\sin^{2m-1}\theta+\frac{E_{0,n;m}}{2m+1}\sin^{2m+1}\theta+\frac{3E_{0,n;m}}{2m}sin^{2m-1}\theta
cos\theta\nonumber\\
&&+\sum_{p=2}^{\left[\frac{n}{2}\right]+1}\frac{4g_{n,p}-4h_{n,p}}{2m+2p-3}sin^{2m+2p-3}\theta+\sum_{p=2}^{\left[\frac{n}{2}\right]+1}\frac{h_{n,p}-3g_{n,p}}{2m+2p-1}sin^{2m+2p-1}\theta\nonumber\\
&&-\sum_{p=2}^{\left[\frac{n}{2}\right]+1}i_{n,p}sin^{2m+2p-3}\theta\cos\theta+\sum_{p=2}^{\left[\frac{n}{2}\right]+1}\frac{g_{n,p}}{2m+2p}sin^{2m+2p-1}\theta\cos\theta\nonumber\\
&&-\sum_{p=l+1}^{\left[\frac{n}{2}\right]+1}j_{n,p}\cos\theta\sum_{l=1}^{p-1}\frac{\bar{I}(2m+2p-4,p-l-1)}{2m+2p-3}\sin^{2m+2l-3}\theta
\end{eqnarray}
It is easy to see the factor $(1-\cos\theta)^3$ in the integrand in
above calculation make the form of $A_{n}(\theta)$ very complicated.
More calculation are need to treat  the problem now. The results are
\begin{eqnarray}
W_n(\theta)&=&\cos\theta
\sum_{l=1}^{[\frac{n}{2}]}a_{n,l}sin^{2l-1}\theta+\sum_{l=1}^{[\frac{n+1}{2}]}b_{n,l}sin^{2l-1}\theta
\end{eqnarray}
\begin{eqnarray}
&&b_{n,l}\nonumber\\
&=&\sum_{p=l+1}^{\left[\frac{n}{2}\right]+1}\frac{3(2m+2l-2)j_{n,p}\bar{I}(2m+2p-4,p-l-1)}{(2m+2l+1)(2m+2l+3)(2m+2p-3)}+\frac{g_{n,l}}{2m+2l}
\xi_{l,1} ,l\ge1 \label{bl}
\end{eqnarray}
\begin{eqnarray}
&&a_{n,l}\nonumber\\
&=&-\sum_{p=l+1}^{\left[\frac{n}{2}\right]+1}\frac{(2m+2l-2)(2m+2l)j_{n,p}\bar{I}(2m+2p-4,p-l-1)}{(2m+2l+1)(2m+2l+3)(2m+2p-3)},l\ge1
\label{al}
\end{eqnarray}
where $\xi_{l,1}=0,\, l=1$ and otherwise $\xi_{l,1}=1, \,l\ne 1$. We
delay the  process of proved in the appendix-1. Therefore, we have
proved the correctness of our induction about the general formula
with $W_n$.That is to say, any $n\geq1$, $W_n$ satisfy the form of
general formula. Interestingly, the general formula with $W_n$ in
the case of $s=3/2$ is same as $s=1/2$.


\begin{thebibliography}{99}
\bibitem{Teu1} Teukolsky S.A., \textit{Phys.Rev.Lett.} \textbf{29}, 1114 (1972).
\bibitem{Teu2} Teukolsky S.A., \textit{Astrophys.J.} \textbf{185}, 635 (1973).
\bibitem{Flammer3} Flammer C 1956 \textit{Spheroidal wave functions}.(Stanford, CA£ºStandford Univiersity Press)
\bibitem{Stratton4} Stratton J 1956 \textit{Spheroidal Wave Functions}. (New York£ºWiley)
\bibitem{Li5}Li L W, Kang X K, Leong M S 2002 \textit{Spheroidal wave functions in electromagnetic theory} ( New York£ºJohn Wiley and Sons Inc.)
\bibitem{Tian6}Tian G H 2005 \textit{Chin.Phys.Lett.}\textbf{22} 3013
\bibitem{Tian7}Tian Guihua, Shuquan Zhong, Solve spheroidal wave functions by SUSY method, arXiv: 0906.4685v2 [gr-qc] 30 Jun 2009.
\bibitem{Tian8}Tian Guihua, Shuquan Zhong, New investigation for the spheroidal wave functions, arXiv: 0906.4687v2 [gr-qc] 30 Jun 2009.
\bibitem{Infeld9} Infeld L, Hull T E 1951 \textit{Rev.Mod.Phys.}\textbf{21} 23
\bibitem{Cooper11}Cooper F, Khare A, Sukhatme U 1995
\bibitem{Tian12} Tian Guihua and Zhong Shuquan 2010 Chin. Phys. Lett. \textbf{27} 040305.
\bibitem{Berti13}Emanuele Berti, Vitor Cardoso and Marc Casals, Eigenvalues
and eigenfunctions of spin-weighted spheroidal harmonics in four and
higher dimensions, arXiv:gr-qc/0511111v4 1 May 2006.
\bibitem{grad} Gradsbteyn I.S., Ryzbik L.M. 2000 Table of integrals,
series, and products. 6th edition, Elsevier(Singapore)pte. Ltd.
\end{thebibliography}
\end{document}